\documentclass[conference]{IEEEtran}
\IEEEoverridecommandlockouts
\usepackage{cite}
\usepackage{amsmath,amssymb,amsfonts}
\usepackage{algorithmic}
\usepackage{graphicx}
\usepackage{textcomp}
\usepackage{xcolor}
\def\BibTeX{{\rm B\kern-.05em{\sc i\kern-.025em b}\kern-.08em
    T\kern-.1667em\lower.7ex\hbox{E}\kern-.125emX}}
\usepackage[ruled, linesnumbered]{algorithm2e}
\usepackage{algorithm2e,setspace}
\usepackage{booktabs}
\usepackage{multirow}

\usepackage{array}
\newcolumntype{L}[1]{>{\raggedright\let\newline\\\arraybackslash\hspace{0pt}}m{#1}}
\newcolumntype{C}[1]{>{\centering\let\newline\\\arraybackslash\hspace{0pt}}m{#1}}
\newcolumntype{R}[1]{>{\raggedleft\let\newline\\\arraybackslash\hspace{0pt}}m{#1}}

\usepackage{mathtools}
\usepackage{amssymb}

\newcommand{\believes}{\mid\equiv}
\newcommand{\sees}{\triangleleft}
\newcommand{\oncesaid}{\mid\sim}

\newcommand{\fresh}[1]{\#(#1)}

\newcommand{\sharekey}[1]{\xleftrightarrow{#1}}

\makeatletter
 \let\old@ps@headings\ps@headings
 \let\old@ps@IEEEtitlepagestyle\ps@IEEEtitlepagestyle
 \def\confheader#1{%
 \def\ps@IEEEtitlepagestyle{%
 \old@ps@IEEEtitlepagestyle%
 \def\@oddhead{\strut\hfill#1\hfill\strut}%
 \def\@evenhead{\strut\hfill#1\hfill\strut}%
 }%
 \ps@headings%
 }
\makeatother

\confheader{%
\parbox{18cm}{\centering Accepted copy for Publication at the 8th IEEE International Conference on Smart and Sustainable Technologies, 2023\\
Final published version available at: https://doi.org/10.23919/SpliTech58164.2023.10193573}}

\begin{document}

\title{Wireless BMS Architecture for Secure Readout in Vehicle and Second life Applications}

\author{
\IEEEauthorblockN{Fikret Basic, Claudia Rosina Laube, Patrick Stratznig, Christian Steger}
\IEEEauthorblockA{\textit{Institute of Technical Informatics} \\
\textit{Graz University of Technology}\\
Graz, Austria \\
\{basic, steger\}@tugraz.at}
\and
\IEEEauthorblockN{Robert Kofler}
\IEEEauthorblockA{\textit{R\&D Battery Management Systems} \\
\textit{NXP Semiconductors Austria GmbH Co \& KG}\\
Gratkorn, Austria \\
robert.kofler@nxp.com}
}

\maketitle

\begin{abstract}
Battery management systems (BMS) are becoming increasingly important in the modern age, where clean energy awareness is getting more prominent. They are responsible for controlling large battery packs in modern electric vehicles. However, conventional solutions rely only on a wired design, which adds manufacturing cost and complexity. Recent research has considered wireless solutions for the BMS. However, it is still challenging to develop a solution that considers both the active in-vehicle and the external second-life applications. The battery passport initiative aims to keep track of the batteries, both during active and inactive use cases. There is a need to provide a secure design while considering energy and cost-efficient solutions. We aim to fill this gap by proposing a wireless solution based on near-field communication (NFC) that extends previous work and provides a unified architecture for both use cases. To provide protection against common wireless threats, an advanced security analysis is performed, as well as a system design analysis for the wake-up process that reduces the daily power consumption of the stored battery packs from milli- to microwatts.
\end{abstract}


\section{Introduction}
\label{sec:intro}
Battery management systems (BMS) represent one of the most important building blocks of modern electric vehicles (EV). They are responsible for ensuring safe and reliable use of large battery packs \cite{Hu2019}. 
Since they are one the main driving forces of an EV, any problem that occurs with the batteries can directly affect the entire vehicle and also the driver's safety, leading to various hazards such as thermal runaway \cite{hartmann_2018}. 
Modern BMS are deployed in several topologies where different modules are responsible for tasks ranging from battery cell sensor data acquisition to their transmission. 

One challenge with modern EV battery systems is the reuse of used batteries at the end of their life cycle. Even if the batteries no longer meet the needs of a vehicle, they may be able to be used for other applications. This battery state is referred to as second life \cite{second_life_book}. 
The battery's state of health (SoH), state of charge (SoC), and other diagnostic information must be tracked during its lifetime \cite{rui_2018}. A BMS controller would track this information in an active EV use case through its internal communications, but if the batteries are stored in a warehouse, for example, the battery packs may only be accessed through an interface with a battery pack controller (BPC) using an external reader. BMS communication services can therefore be viewed from two angles: (i) internal - for communication between different sensors and bridge modules, and (ii) external - for diagnostic and monitoring purposes read from outside the system. There are several legislations underway aimed to introduce battery tracking via battery passports and even distributed monitoring in the cloud using electronic passports \cite{batt_passport_eu_reg, BERGER2022131492}. It is important to find an affordable system design that allows flexible and fast readout with respect to both communication approaches. However, currently, there is no unified design that considers different BMS topologies. 

Another challenge that is common with BMS is the use of cables for communication. The use of cables is associated with higher assembly cost, weight, complexity, limited scalability, and maintenance \cite{electronics10182193, fabian_2021, taesic_2018}. However, they also provide fast and reliable transmission between modules, which is important for systems such as BMS. An alternative would be to use wireless technologies, but here it is important to assess which technology should be used. There is currently no clear answer to this, with several works proposed ranging from the use of BLE \cite{fabian_2021, cody_2015}, Wi-Fi \cite{tudor_2018, huang_2020}, ZigBee \cite{Rahman_2017}, and other technologies \cite{electronics10182193}. 
Using multiple technologies in one system is not practical and can lead to increased costs and radio interference. Our goal is to present a design that unifies different use cases under the same wireless technology. 
We aim to achieve this through the use of radio-frequency identification (RFID), specifically accessible and low-cost near-field communication (NFC).

The BMS and batteries are vital components in modern EVs that also require adequate protection. Any vulnerable system component should be authenticatable and provide protection against tampering and even eavesdropping to protect user privacy 
\cite{Sripad2017VulnerabilitiesOE, Cheah2019, plosz_2014}. 
NFC offers advantages from a security perspective because it enables short-range communication that is difficult to be exploited remotely. However, there are still many vulnerabilities in BMS and NFC that need to be addressed to provide a fully secure design \cite{plosz_2014, chattha_2014, Haselsteiner2006SecurityIN}. 

In response to the aforementioned challenges, we present in this paper a complete design for wireless BMS internal and external data communication based on NFC. To extend on the previous work in this area, we unify two separate designs, one targeting internal sensor readout from battery packs and the other targeting external status and diagnostic readout in a BMS system design~\cite{basic_journal_2022, basic_mobile_2022}. 

\textbf{Contributions.} We extend the security architecture by presenting and formally verifying a complete authentication and session establishment protocol for the external readout use case. We also investigate the important wake-up readout cycle for other use cases and propose two different system designs.

Summarized, our main contributions are:
\begin{itemize}
    \item A system architecture that enables wireless internal and external readout of battery sensor data for various cases.
    \item Proposal of two methods for the wake-up process. 
    \item Extending the existing security solution and its proof.
\end{itemize}

\section{Background and Related Work}
\subsection{Battery management systems (BMSs)}

A BMS is responsible for controlling a large number of battery cells connected either in series or in parallel to ensure safe and reliable operation \cite{Hu2019}.
There are several BMS topologies, with distributed systems being the most common \cite{andrea_2020}. Here, a powerful main BMS controller is responsible for controlling and relaying information to the outside world. The controller communicates with several BPCs that are responsible for intermediate operations and data collection from a group of battery cells.
We call this group of elements a battery pack.

\subsection{Near field communication (NFC)}

NFC is a short-range wireless technology defined in the ISO 18092 NFC standard, which builds on the ISO 14443 RFID standard. It operates at a radio frequency (RF) of 13.56\,MHz with a range and data rate of up to 10\,cm and 848\,kbit/s, respectively \cite{iso_nfc_standards}. Today, it is used for fast terminal controls, payments, and applications that benefit from energy harvesting. NFC can operate in three main modes, reader/writer mode, peer-to-peer mode, and card emulation mode \cite{iso_nfc_standards}. In our work, we rely on the reader/writer mode for all use cases. Here, devices are divided into an active and a passive class. In the context of BMS, NFC suffers from a short range compared to other wireless technologies~\cite{electronics10182193}, but it provides an advantage when considering interference, security concerns and accessibility with NFC-enabled devices \cite{basic_journal_2022}. It is also capable of generating electromagnetic fields through an active device to power the passive devices via energy harvesting \cite{madjda_2019}. 

\subsection{Second life for batteries}

A common problem with the use of batteries in EVs is the problem of battery recycling. When the maximum battery capacity falls below a certain threshold (around 80\%), the current batteries need to be replaced with new ones. However, the high recycling cost and pollution make this process undesirable. Instead, it is proposed to maximize the potential use of a battery by reusing it for other uses that are not constrained by the same customer criteria as with EVs~\cite{second_life_book}. 

Second-life battery use is an ongoing discussion within the European Union (EU).
A concept of "battery passports" is being discussed, which aims to track batteries throughout their lifetime, from manufacturing to recycling~\cite{batt_passport_eu_reg, BERGER2022131492}. 
The tracking and identification of batteries would be performed via QR codes. However, QR technology is limited in terms of functionality, scalability, services and security. In this work, we propose an alternative lightweight design that aims to fill these drawbacks with minimal additional overhead. 
The use of battery passports is also of interest with respect to the new concepts related to the BMS service extension to the cloud environment, as shown in the work of Li et al.~\cite{digital_twin_2020}, Taesic et al.~\cite{taesic_2018}, and Kai et al.~\cite{li_cloud_2020}. This cloud shift provides additional power for SoC and SoH computations and enables machine learning or digital twin models, and extra security verification.

\section{The Novel BMS Design for Wireless Readout}
\label{sec:design}
\begin{figure}[!t]
  \centering
    \vspace{-1.00mm} 
  \includegraphics[width=0.98\linewidth]{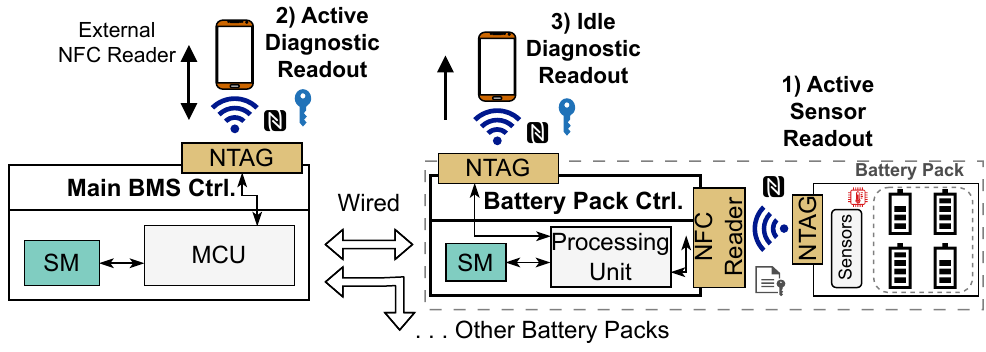}
    \vspace{-2.50mm} 
  \caption{Wireless NFC BMS architecture considering three different use cases, extending basis design with secure modules (SM) and NFC tags and readers.}
  \label{fig:bms_main_arch}
    \vspace{-1.75mm} 
\end{figure}

\begin{table}[!t]
\caption{Overview of the proposed use cases and deployment scenarios.}
    \vspace{-2.50mm} 
\label{table:wireless_bms_use_cases}
\begin{center}
\begin{tabular}{@{}cccc@{}}
\toprule
    \multicolumn{1}{C{1.7cm}}{\textbf{Use case}} & \multicolumn{1}{C{1.8cm}}{\textbf{Deploy. scenario}} & \multicolumn{1}{C{1.8cm}}{\textbf{Comm. direction}} & \multicolumn{1}{C{1.8cm}}{\textbf{Readout data}} \\ \midrule
    \multicolumn{1}{C{1.7cm}}{Active sensor} & \multicolumn{1}{C{1.8cm}}{internal; deployed} & \multicolumn{1}{C{1.8cm}}{Batt. pack $\rightarrow$ BPC} & \multicolumn{1}{C{1.8cm}}{sensor data} \\ \midrule[0.15pt]
    \multicolumn{1}{C{1.7cm}}{Idle diagnostic} & \multicolumn{1}{C{1.8cm}}{external; stored} & \multicolumn{1}{C{1.8cm}}{BPC $\rightarrow$ Ext. reader} & \multicolumn{1}{C{1.8cm}}{status \& sensor data} \\ \midrule[0.15pt]
    \multicolumn{1}{C{1.7cm}}{Active diagnostic} & \multicolumn{1}{C{1.8cm}}{external; deployed} & \multicolumn{1}{C{1.8cm}}{BMS $\leftrightarrow$ Ext. reader} & \multicolumn{1}{C{1.8cm}}{status \& diagnostic} \\
\bottomrule
\end{tabular}
\end{center}
\end{table}

By relying only on NFC, we are able to provide a solution for three different scenarios that extend the BMS functionality, as described in Table~\ref{table:wireless_bms_use_cases}.: (i) active readout of internal sensors~\cite{basic_journal_2022}, (ii) idle state scenario considered for off-vehicle use cases, and (iii) active diagnostic readout~\cite{basic_mobile_2022}.

The system architecture is shown in Figure~\ref{fig:bms_main_arch}. For both diagnostic readouts (active and idle), both devices are first authenticated before a secure channel is established. The data link communication layer is handled with NDEF records. 
The data storage depends on the reader application, either storing it offline or online in a database as proposed for battery passports~\cite{batt_passport_eu_reg}.
Elements of the design and its implementation were handled in a recent master’s thesis~\cite{mastersthesis}. 

\begin{figure}[!t]
  \centering
    \vspace{-1.00mm} 
  \includegraphics[width=0.95\linewidth]{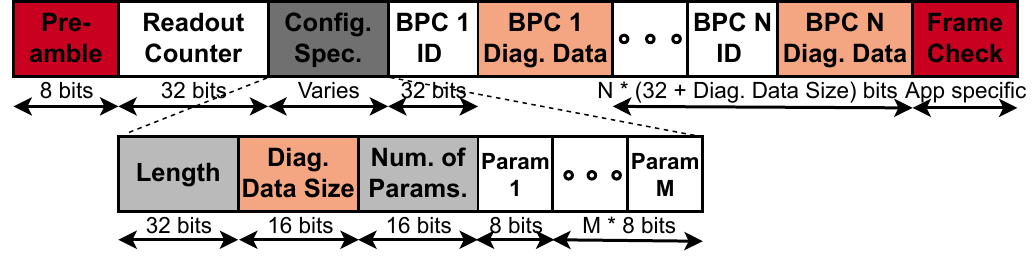}
    \vspace{-2.50mm} 
  \caption{Application packet structure for the active diagnostic use case.}
  \label{fig:diag_data_struct}
    \vspace{-1.75mm} 
\end{figure}

Diagnostic readout for active and idle states differentiate on the devices used and the data direction.
In both cases, communication takes place with an external NFC reader. The active diagnostics use case is intended for situations where BMSs are already deployed in a neighbouring system. In this case, the BMS controller is able to provide the current diagnostic data. The idle readout is intended for the storage of battery packs during their second life transfer~\cite{second_life_book}. Here, it is considered beneficial to occasionally check the batteries' SoH to determine that no hazards have occurred. In the case of active diagnostic readout, the application data is first formatted before being transferred from the BMS host controller to the NFC reader. The proposed lightweight structure is shown in Figure~\ref{fig:diag_data_struct}. The BMS is expected to first collect the data from each BPC before forwarding it to the NFC reader. 
However, BMS topologies need to also be considered.
The following considerations must be made for other topologies~\cite{andrea_2020}:
\begin{itemize}
    \item Centralized: consists of one main BMS control unit and no intermediate modules. Requires one NTAG and NFC reader. The idle diagnostic use case is considered unfeasible unless the controller is also stored.
    \item Modulated: multiple modules, with one main module and multiple followers. The main module enables the external interface, but each follower module also needs to include an NTAG and NFC reader for all use cases.
    \item Distributed: similar to the modulated design, but with a clearer separation between the central BMS controller and the battery packs. The architecture is shown in Figure~\ref{fig:bms_main_arch}.
    \item Decentralized: consists of multiple BMS subsystems. Considers linear growth of required NFC interfaces based on the topology of the underlying BMS subsystem.
\end{itemize}

\begin{figure}[!t]
  \centering
    \vspace{-1.00mm} 
  \includegraphics[width=0.95\linewidth]{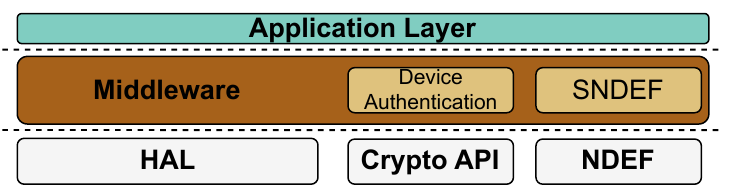}
    \vspace{-2.50mm} 
  \caption{Proposed wireless BMS software layer stack.}
  \label{fig:sw_arch}
    \vspace{-1.75mm} 
\end{figure}

\textbf{Software architecture.} The software architecture is generalized and is shown in Figure~\ref{fig:sw_arch}. It consists of three main layers. At the bottom is the hardware abstraction layer (HAL). It is vendor-specific and contains the driver components, e.g., for interfaces, clock, etc. The cryptographic application programming interface (API) is considered an independent entity that can be connected either directly to the HAL or to a separate security module. NDEF messages are controlled by the NFC driver. Their payload contains the components of the higher SNDEF data set located at the middle layer, i.e., the middleware.
Independent of the SNDEF, the middleware is also responsible for device authentication. This means that no other access to the application layer is possible unless it has first been verified, confirmed, and processed by the middleware.
Since the middle layer imposes no restrictions on the data format, the application layer can be freely defined by the developer.

\section{Security Model}
\label{sec:security_model}
To protect against common threats, a security model needs to be introduced for both internal and external communication. Several BMS threat analysis models have been created, each highlighting important vulnerabilities~\cite{Sripad2017VulnerabilitiesOE, Cheah2019, taesic_2018}.
We aim not to invent a new protocol that could be susceptible to untracked attacks, but rather to adapt simple, yet effective solutions. The security model needs to answer to the current known BMS and NFC threats and fulfil the following requirements:
\begin{itemize}
    \item Mutual device authentication
    \item Use of secure channels: encryption and tamper-proof
    \item Lightweight models with minimal performance overhead
\end{itemize}

\subsection{Securing battery pack internal readout}
\label{lab:sensor_readout}

Because battery packs are enclosed, any form of physical attack would directly damage the components and would therefore be infeasible or too costly for the attackers. Similarly, an external attack would also likely be difficult or impossible, although further research is needed to confirm these claims. This leaves the protection of reading the battery packs based on the proposal by Basic et al. in \cite{basic_journal_2022} sufficient, i.e., authentication protection based on either symmetric or asymmetric cryptography with previously embedded keys.
Hence, on the device-level protection, two key strategies can be employed: (i) authentication; symmetric with pre-shared keys, or asymmetric with originality signatures, and (ii) authorization, with password-controlled read and write access.

\subsection{Security protocol for external BMS readout}
\label{lab:security_protocol}

The protocol is based on symmetric, rather than asymmetric cryptography, to account for the potential performance or hardware limitations.
If the master key is leaked, all past and future sessions could be compromised. Therefore, it is of utmost importance to secure the location of the master key with an SM. We rely on the use of SNDEF records~\cite{basic_mobile_2022}. 

\begin{figure}[!t]
  \centering
    \vspace{-1.00mm} 
  \includegraphics[width=0.95\linewidth]{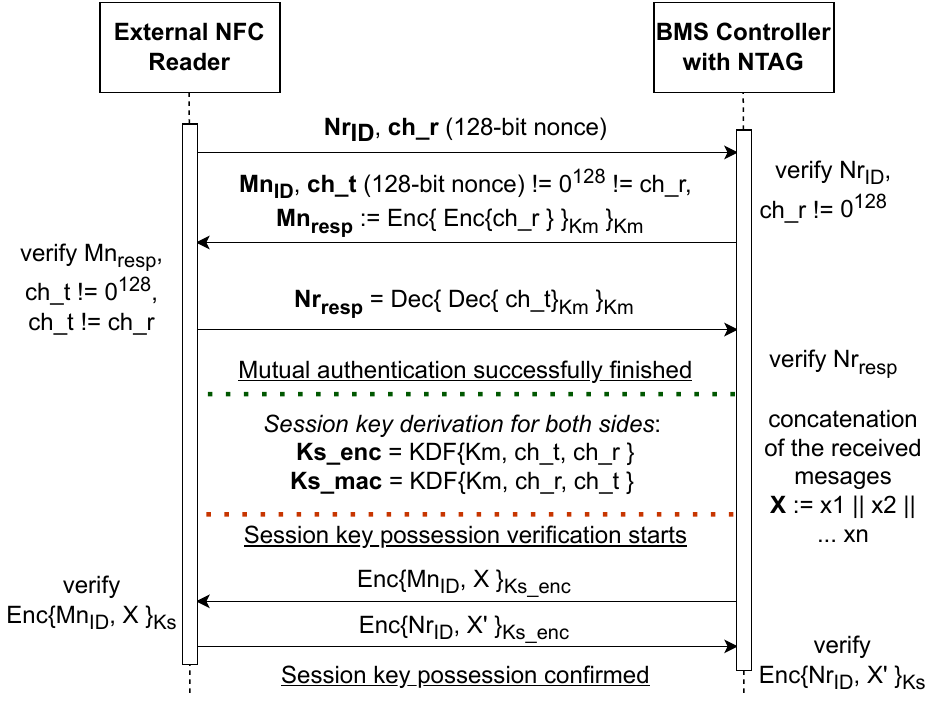}
    \vspace{-2.50mm} 
  \caption{Security protocol sequence diagram for the external BMS readout.}
  \label{fig:bms_nfc_ext_sec_protocol}
    \vspace{-1.75mm} 
\end{figure}

The security design for external readout, which considers the idle state and active diagnostics use cases, is based on the lightweight security design for authentication and data exchange proposed in \cite{basic_mobile_2022}. We formalize the protocol in Figure~\ref{fig:bms_nfc_ext_sec_protocol}. and extend the solution to consider some additional security vulnerabilities that may arise from the original design. It consists of two phases, the authentication phase and the newly added session key possession confirmation phase. During the session key confirmation phase, previous messages are appended
, to confirm session key ownership for both parties.
\begin{align}
    & 1)\;N_R \rightarrow M_N : N_R, ch_r  \\
    & 2)\;M_N \rightarrow N_R : M_N, ch_t, \{\{M_N, ch_r\}_{K_M}\}_{K_M} \\
    & 3)\;N_R \rightarrow M_N : \{\{N_R, ch_t\}_{K_M}^{-1}\}_{K_M}^{-1} \\
    & 4)\;M_N \rightarrow N_R : \{M_N, X\}_{K_S} \\
    & 5)\;N_R \rightarrow M_N : \{N_R, X'\}_{K_S}
\end{align}

From the protocol; $N_R$: NFC reader id., $M_N$: BMS ctrl. id., $ch_r$: challenge nonce request from $N_R$, $ch_t$: challenge nonce request from $M_N$, $K_M$: pre-embedded master key, $K_S$: session key, $X$: concatenated previously received messages from $N_R$, $X'$: concatenated received messages from $M_N$.

The double key encryption is used to fend off an oracle attack that could expose the vulnerabilities of the CMAC if used in the key derivation function (KDF) when based on the CBC-MAC computation. In addition, a nonce cannot be zero or equal to another. 
Another important point is that the underlying ciphers are not identical, i.e., in our case, BPC encrypts, while the external NFC reader decrypts the challenges, to protect against ``chosen challenge oracle'' attacks.

\section{Wake Up Model Design}
\label{sec:wakeup_model}
\begin{figure*}[!t]
  \centering
    \vspace{-1.00mm} 
  \includegraphics[width=0.96\linewidth]{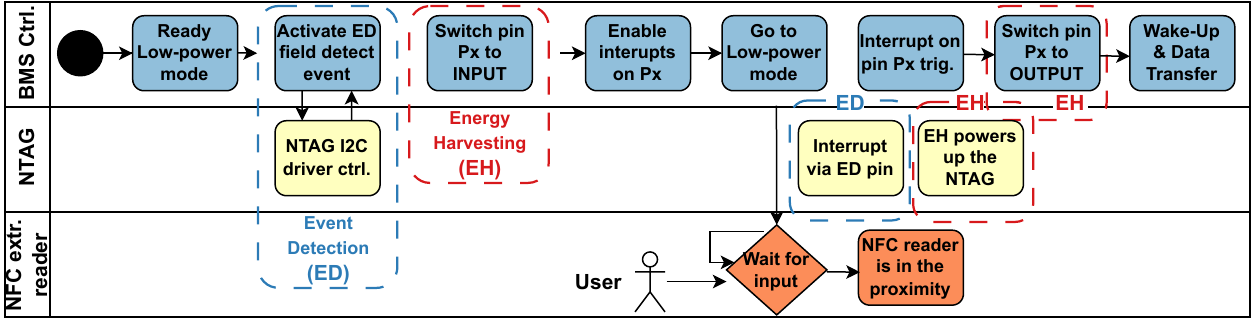}
    \vspace{-2.50mm} 
  \caption{Wake-up process flowchart: in dashed (ED) blocks steps are shown for the event detection, whereas in (EH) they are shown for the energy harvesting.}
  \label{fig:ed_eh_flowchart}
    \vspace{-1.75mm} 
\end{figure*}

Previous works mainly investigated system design and communication with NFC. There are no clear design specifications for interaction with BMS controllers. 
The wake-up application is important for the idle state use case. Since a BPC is stored along with battery cells, it is not necessary to keep the controller on during the entire storage period. By controlling the wake-up of the BPC, it would be possible to minimize the total power consumption during the time the battery cells are stored and provide power during a communication period with an external NFC reader. Therefore, we want to propose a design that meets the following two main requirements:
\begin{enumerate}
    \item Low power consumption through mode switching 
    \item Fast wake-up time
\end{enumerate}

We propose two approaches to wake-up system design based on the different characteristics of BPC and NTAG: (i) with \textit{event detection} (ED) and (ii) with \textit{energy harvesting} (EH). They are evaluated based on total wake-up time and power consumption, as well as design requirements and HW/SW design considerations. Figure~\ref{fig:ed_eh_flowchart}. shows the flowchart, with separate steps for both designs. Both rely on the use of $I^2C$ for data transfer, but the difference is in how the wake-up events are triggered and how the power states are managed. 

\textbf{Wake-up with event detection.} This method uses the event detection (ED) function on modern NTAG boards, where the ED pin acts as an event pin that can respond to different events, specifically here, the presence of an NFC field. When an RF field is detected, the ED pin is set to logic high. During idle time, the NTAG remains in the standby state and is constantly powered by the host BPC. The lowest theoretical consumption that can be achieved is if the host BPC uses a very low power state (VLPS) that can still power the NTAG and respond with the wake-up, with the NTAG being in standby mode.

\textbf{Wake-up with energy harvesting.} In this method, the NTAG can be completely shut down and the BPC can be put into a VLPS mode, as with the ED method. This results in lower power consumption, but may also result in a slightly longer wake-up time since the board wake-up depends on energy harvesting. The NTAG is put into EH mode, where energy is harvested from the RF field in close contact with the antenna. Since the BPC does not power the NTAG at all during the sleep phase, unlike the ED method, the NTAG must remain powered on after energy harvesting has triggered its wake-up. Thus, after the BCP wakes up and during the open communication session, it draws power from the battery cells for its normal operating mode and now also powers the NTAG.

\begin{table}[!t]
\renewcommand{\arraystretch}{1.00}
\setstretch{0.90} 
\caption{Overview of the proposed wake-up system design approaches.}
    \vspace{-1.50mm} 
\label{table:wakeup_system_design}
\begin{center}
\begin{tabular}{@{}c>{\raggedright\arraybackslash}m{2.8cm}>{\raggedright\arraybackslash}m{3.3cm}@{}}
\toprule
    \multicolumn{1}{C{1.5cm}}{\textbf{Model}} & \multicolumn{1}{C{2.8cm}}{\textbf{Prerequisites}} & \multicolumn{1}{C{3.3cm}}{\textbf{Pros / Cons}} \\ \midrule
    \multicolumn{1}{C{1.5cm}}{Event Detection (ED)} & 
    NTAG needs to have an event pin; requires a constant power source
    & 
    [+] wake-up is possibly faster, [-] needs constant power source, [-] higher power consumption
    \\
    \midrule[0.15pt]
    \multicolumn{1}{C{1.5cm}}{Energy Harvest. (EH)} & 
    NTAG’s EH needs to be specially configured; the reader has to have EH enabled
    & 
    [+] NTAG is powered off in idle, [+] after wake-up, BPC can supply the NTAG, [-] wake-up takes longer
    \\
\bottomrule
\end{tabular}
\renewcommand{\arraystretch}{1.00}
\setstretch{1.00} 
\vspace{-1.75mm} 
\end{center}
\end{table}

\section{Evaluation}
For the evaluation, we use the NCx33xx series of products from NXP as the NTAG and NFC Reader. They are NFC forum-compliant automotive-graded components that provide several benefiting features for testing~\cite{ntag_descr}.
The S32K144 microcontroller board was chosen as both the main BMS controller and the BPC. It is widely used in automotive and industrial applications, and it offers a ready BMS diagnostic application~\cite{s32_desc}.
We evaluate the design on the security aspects both with informal and formal protocol analysis and on the real-device performance analysis by investigating the application overhead of the wake-up process. 

\subsection{Security informal analysis}

The goal is to protect three main assets: (A1) hardware integrity, (A2) software integrity, (A3) transmitted data. To protect the integrity of (A2) and (A3), the protocol uses AES as the encryption algorithm and CMAC for tag verification. This is done to increase usability on different devices and improve performance. The operation mode is CBC in Encrypt-then-MAC mode, as it provides higher security than other operation modes.
It is important to use a different key between the AES and CMAC computation because using the same key would allow the attacker to forge the tags if they had access to an encryption oracle where they could query the values of the last CMAC computation block.
Additionally, to protect against forms of replay attack, tag chaining was implemented. Tag chaining considers appending the previous tag into the buffer of the newly received message to calculate the total tag value, i.e., the MAC value, by using the following structure:
\begin{equation*}
    MAC_{buffer} := sec\_data\; | \;IV\; | \;add\_data\; | \;previous\_tag
\end{equation*}
(A1) is guaranteed by performing device authentication on the battery pack as described in Section~\ref{lab:sensor_readout}, and by mutual authentication when communicating with an NFC reader. 

\subsection{Security formal analysis - BAN logic}

A formal protocol analysis was done on the mutual authentication and session establishment protocol presented in Section~\ref{lab:security_protocol}. It uses the BAN logic and its postulates \cite{michael_1990}.

\textit{Idealized protocol.}
We use the protocol from Section~\ref{lab:security_protocol}:
\begin{align}
    & 1)\;all\;\;plaintext \\
    & 2)\;M_N \rightarrow N_R : \{\{ch_r, N_R \sharekey{K_M} M_N \}_{K_M}\}_{K_M} \\
    & 3)\;N_R \rightarrow M_N : \{\{ch_t, N_R \sharekey{K_M} M_N \}_{K_M}\}_{K_M} \\
    & 4)\;M_N \rightarrow N_R : \{X, N_R \sharekey{K_S} M_N \}_{K_S} \\
    & 5)\;N_R \rightarrow M_N : \{X', N_R \sharekey{K_S} M_N \}_{K_S}
\end{align}
\textit{Initial assumptions.} The following assumptions are made. Firstly, both devices regard the sent nonces to be fresh:
\begin{equation}
    N_R \believes \fresh{ch_r} \;,\; M_N \believes \fresh{ch_t}
\end{equation}
Both sides believe in the use of the shared master key:
\begin{equation}
    N_R \believes N_R \sharekey{K_M} M_N \;,\; M_N \believes N_R \sharekey{K_M} M_N
\end{equation}
\textit{Goals.} We want to make sure that both parties are mutually authenticated and know that the other side trusts that as well: 
\begin{align}
    & G1.1)\;\; N_R \believes M_N \believes N_R \sharekey{K_M} M_N \\
    & G1.2)\;\; M_N \believes N_R \believes N_R \sharekey{K_M} M_N
\end{align}
For the second-order goals, we want to make sure that both parties trust that the other party has the correct session key:
\begin{align}
    & G2.1)\;\; N_R \believes M_N \believes N_R \sharekey{K_S} M_N \\
    & G2.2)\;\; M_N \believes N_R \believes N_R \sharekey{K_S} M_N
\end{align}
\textit{Verification.} We will start first with G1.1 and G1.2 goals. To verify, we will apply the rules described in the BAN logic~\cite{michael_1990}. 
On Eq.~7 we first apply the \textit{message-meaning rule}:
\begin{equation}
\label{eq:message_meaning_rule_1}
    \frac{N_R \believes N_R \sharekey{K_M} M_N, N_R \sees \{\{ch_r\}_{K_M}\}_{K_M} }{N_R \believes M_N \oncesaid (ch_r,\; N_R \sharekey{K_M} M_N) }
\end{equation}
We use the \textit{freshness rule} for the nonce and key statement:
\begin{equation}
\label{eq:fresh_nonce_1}
    \frac{\fresh{ch_r} }{\fresh{ch_r, N_R \sharekey{K_M} M_N}}
\end{equation}
Then, on (\ref{eq:fresh_nonce_1}) \& (\ref{eq:message_meaning_rule_1}) we can apply the \textit{nonce verification rule}:
\begin{align}
    \frac{N_R \believes (\ref{eq:fresh_nonce_1}), N_R \believes M_N \oncesaid (ch_r, N_R \sharekey{K_M} M_N) }{N_R \believes M_N \believes (ch_r,\; N_R \sharekey{K_M} M_N)}
\end{align}
Finally, to verify the goal G1.1, we can take the \textit{belief rule}:
\begin{equation}
    \frac{N_R \believes M_N \believes (ch_r,\; N_R \sharekey{K_M} M_N)}{N_R \believes M_N \believes N_R \sharekey{K_M} M_N}
\end{equation}
The verification of goal G1.2 is symmetrical to G1.1, and thus also proved. For the second-order goals, we set additional assumptions. Since $X,\; X'$ from (9) and (10) are composed out of $ch_r$ and $ch_t$, we assume by \textit{freshness rule} that:
\begin{equation}
    \frac{N_R \believes \fresh{ch_r} }{N_R \believes \fresh{X} } \;,\; \frac{M_N \believes \fresh{ch_t} }{M_N \believes \fresh{X'} }
\end{equation}
Now, using the \textit{belief rule} we get the important statements that both sides believe the session key possession:
\begin{equation}
    \frac{N_R \believes (X,\; N_R \sharekey{K_S} M_N) }{N_R \believes N_R \sharekey{K_S} M_N}
\end{equation}
\begin{equation}
    \frac{M_N \believes (X',\; N_R \sharekey{K_S} M_N) }{M_N \believes N_R \sharekey{K_S} M_N}
\end{equation}
To finalize the verification of G2.1 \& G2.2, we use the same line of rules as for the G1.1 \& G1.2, i.e., by applying the message-meaning rule, then the freshness rule, nonce-verification and finally the belief rule. Here, G2.2 is also symmetrical in verification to G2.1.


\subsection{Wake-up process evaluation}

\begin{figure}[!t]
  \centering
  \vspace{-1.00mm} 
  \includegraphics[width=0.95\linewidth]{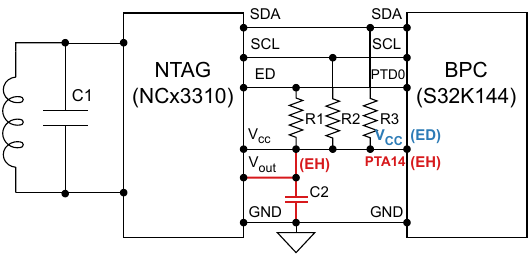}
  \vspace{-2.50mm} 
  \caption{Hardware design for the BPC that considers the wake-up use cases.}
  \label{fig:ed_eh_circuit}
  \vspace{-1.75mm} 
\end{figure}

A real-world implementation was made using the components based on the hardware design presented in Figure~\ref{fig:ed_eh_circuit}. To test the design of ED, the NCx3310 is placed in standby mode with the S32K144 using the VLPS mode. For the idle phase, this results in $29.8\,\mu A$ from the S32K144 and $5.9\,\mu A$ from the NCx3310, for a total of $35.7\,\mu A$ as theoretical current consumption and power consumption of $117.81\,\mu W$. In the EH design case, the NTAG is completely disabled, which means that all power consumption comes only from the BPC, resulting in a theoretical current consumption of $29.8\,\mu A$ and power consumption of $98.3\,\mu W$ for our devices.
While the design shows a working and usable test implementation, in a real application a level shifter should be considered to compensate for the potential cross currents that can occur on the connections between I2C, the return signal port, $V_{cc}$ and $V_{out}$.
Depending on the use case, both methods can be applied, but if the overall goal is to reduce power consumption, the method with EH is proposed. In this mode, the BPC remains in power-saving mode while the NTAG draws no power because it is completely powered down.

\section{Conclusion}
In this work, we have presented a novel design for the wireless deployment of BMS using NFC as an enabling technology, extending previous work. We have considered both internal and external system readouts, as well as active and inactive use cases. We show that it is feasible to use the BMS with NFC for both current active in-vehicle applications and second-life applications. An NFC interface is useful here as it allows flexible readout of the stored battery packs. Here, the security model was extended and evaluated using both informal security analysis and BAN logic. In addition, two wake-up system designs have been proposed and evaluated.
For future work, further investigation of the wake-up method is planned, as well as investigation of other potential RFID options focused on lightweight applications, such as a fast readout of the ID of the stored battery packs.

\section*{Acknowledgment}
This paper is supported by the OPEVA project that has received funding within the Key Digital Technologies Joint Undertaking (KDT JU) from the European Union’s Horizon Europe Programme and the National Authorities (France, Belgium, Czechia, Italy, Portugal, Turkey, Switzerland), under grant agreement 101097267.
Views and opinions expressed are however those of the author(s) only and do not necessarily reflect those of the European Union or KDT JU. Neither the European Union nor the granting authority can be held responsible for them.

\bibliographystyle{ieeetr}
\bibliography{references}

\begin{thebibliography}{10}

\bibitem{Hu2019}
X.~Hu {\em et~al.}, ``{State estimation for advanced battery management: Key
  challenges and future trends},'' {\em Renew. \& Sustain. Energy Reviews},
  2019.

\bibitem{hartmann_2018}
M.~Hartmann and J.~Kelly, ``{Thermal Runaway Prevention of Li-ion Batteries by
  Novel Thermal Management System},'' in {\em IEEE ITEC}, 2018.

\bibitem{second_life_book}
W.-C. Lih, J.-H. Yen, F.-H. Shieh, and Y.-M. Liao, ``{Second Use of Retired
  Lithium-ion Battery Packs from Electric Vehicles: Technological Challenges,
  Cost Analysis and Optimal Business Model},'' in {\em IS3C}, 2012.

\bibitem{rui_2018}
R.~Xiong {\em et~al.}, ``{Critical Review on the Battery State of Charge
  Estimation Methods for Electric Vehicles},'' {\em IEEE Access}, vol.~6, 2018.

\bibitem{batt_passport_eu_reg}
``{Proposal for a regulation of the European Parliament and of the Council
  concerning batteries and waste batteries, repealing Directive 2006/66/EC and
  amending Regulation (EU) No 2019/1020}.''
  https://eur-lex.europa.eu/legal-content/EN/TXT/?uri=CELEX:52020PC0798, 2020.

\bibitem{BERGER2022131492}
K.~Berger, J.-P. Schöggl, and R.~J. Baumgartner, ``Digital battery passports
  to enable circular and sustainable value chains: Conceptualization and use
  cases,'' {\em Journal of Cleaner Production}, vol.~353, 2022.

\bibitem{electronics10182193}
A.~Samanta and S.~S. Williamson, ``{A Survey of Wireless Battery Management
  System: Topology, Emerging Trends, and Challenges},'' {\em Electronics},
  vol.~10, no.~18, 2021.

\bibitem{fabian_2021}
F.~A. Rincon~Vija {\em et~al.}, ``{From Wired to Wireless BMS in Electric
  Vehicles},'' in {\em 17th MSN}, pp.~255--262, 2021.

\bibitem{taesic_2018}
T.~Kim {\em et~al.}, ``{Cloud-Based Battery Condition Monitoring and Fault
  Diagnosis Platform for Large-Scale Lithium-Ion Battery Energy Storage
  Systems},'' {\em Energies}, vol.~11, no.~1, 2018.

\bibitem{cody_2015}
C.~Shell {\em et~al.}, ``{Implementation of a Wireless Battery Management
  System (WBMS)},'' in {\em IEEE I2MTC}, pp.~1954--1959, 2015.

\bibitem{tudor_2018}
T.~Gherman {\em et~al.}, ``{Smart Integrated Charger with Wireless BMS for
  EVs},'' in {\em 44th IECON}, pp.~2151--2156, 2018.

\bibitem{huang_2020}
X.~Huang {\em et~al.}, ``{Wireless Smart Battery Management System for Electric
  Vehicles},'' in {\em IEEE ECCE}, pp.~5620--5625, 2020.

\bibitem{Rahman_2017}
A.~Rahman, M.~Rahman, and M.~Rashid, ``{Wireless Battery Management System of
  Electric Transport},'' {\em {IOP} Conference Series: Materials Science and
  Engineering}, vol.~260, nov 2017.

\bibitem{Sripad2017VulnerabilitiesOE}
S.~Sripad {\em et~al.}, ``{Vulnerabilities of Electric Vehicle Battery Packs to
  Cyberattacks on Auxiliary Components},'' {\em ArXiv}, vol.~1711.04822, 2017.

\bibitem{Cheah2019}
M.~Cheah and R.~Stoker, ``{Cybersecurity of Battery Management Systems},'' {\em
  HM TR series}, vol.~10, no.~3, p.~8, 2019.

\bibitem{plosz_2014}
S.~Plosz {\em et~al.}, ``{Security Vulnerabilities and Risks in Industrial
  Usage of Wireless Communication},'' in {\em IEEE ETFA}, pp.~1--8, 2014.

\bibitem{chattha_2014}
N.~A. Chattha, ``{NFC — Vulnerabilities and Defense},'' in {\em CIACS}, 2014.

\bibitem{Haselsteiner2006SecurityIN}
E.~Haselsteiner and K.~Breitfuss, ``{Security in Near Field Communication (NFC)
  Strengths and Weaknesses},'' in {\em Workshop on RFID Security}, 2006.

\bibitem{basic_journal_2022}
F.~Basic, M.~Gaertner, and C.~Steger, ``{Secure and Trustworthy NFC-Based
  Sensor Readout for Battery Packs in Battery Management Systems},'' {\em IEEE
  Journal of Radio Frequency Identification}, vol.~6, 2022.

\bibitem{basic_mobile_2022}
F.~Basic {\em et~al.}, ``{A Novel Secure NFC-based Approach for BMS Monitoring
  and Diagnostic Readout},'' in {\em IEEE RFID}, pp.~23--28, 2022.

\bibitem{andrea_2020}
A.~Reindl, H.~Meier, and M.~Niemetz, ``{Scalable, Decentralized Battery
  Management System Based on Self-organizing Nodes},'' in {\em ARCS}, 2020.

\bibitem{iso_nfc_standards}
``{Standards - ISO 14443, ISO 15693, NFC-Forum und NDEF}.''
  https://www.nfc-tag-shop.de/info/nfc-entwicklung/standards-iso-14443-nfc-forum-und-ndef.html,
  2017.
\newblock Accessed: 16.01.2023.

\bibitem{madjda_2019}
M.~Bouklachi {\em et~al.}, ``{Energy Harvesting of a NFC Flexible Patch for
  Medical Applications},'' in {\em IEEE WPTC}, pp.~249--252, 2019.

\bibitem{digital_twin_2020}
W.~Li {\em et~al.}, ``Digital twin for battery systems: Cloud battery
  management system with online state-of-charge and state-of-health
  estimation,'' {\em Journal of Energy Storage}, vol.~30, 2020.

\bibitem{li_cloud_2020}
K.~Li {\em et~al.}, ``Battery life estimation based on cloud data for electric
  vehicles,'' {\em Journal of Power Sources}, vol.~468, p.~228192, 2020.

\bibitem{mastersthesis}
C.~Laube, ``Design and implementation of secure {NFC}-based logging for
  stationary battery management systems,'' Master's thesis, TUG, 2022.

\bibitem{ntag_descr}
{NXP Semiconductors}, ``{Ntag 5 link - NFC forum-compliant I2C bridge}.''
  NTP53x2, Rev. 3.3, 2020.

\bibitem{s32_desc}
{NXP Semiconductors}, ``{S32k1xx data sheet}.'' S32K1XX, Rev. 14, 2021.

\bibitem{michael_1990}
M.~Burrows, M.~Abadi, and R.~Needham, ``{A Logic of Authentication},'' {\em ACM
  Trans. Comput. Syst.}, vol.~8, p.~18–36, feb 1990.

\end{thebibliography}

\end{document}